\documentclass{article}

\usepackage{PRIMEarxiv}

\usepackage[utf8]{inputenc} % allow utf-8 input
\usepackage[T1]{fontenc}    % use 8-bit T1 fonts
\usepackage{hyperref}       % hyperlinks
\usepackage{url}            % simple URL typesetting
\usepackage{booktabs}       % professional-quality tables
\usepackage{amsfonts}       % blackboard math symbols
\usepackage{nicefrac}       % compact symbols for 1/2, etc.
\usepackage{microtype}      % microtypography
\usepackage{lipsum}
\usepackage{fancyhdr}       % header
\usepackage{graphicx}       % graphics
\graphicspath{{media/}}     % organize your images and other figures under media/ folder

\usepackage{graphicx}
\usepackage{subcaption}
\newcommand{\quotes}[1]{``#1''}
\newcommand{\absdiv}[1]{%
  \par\addvspace{.5\baselineskip}% adjust to suit
  \noindent\textbf{#1}\quad\ignorespaces
}

\title{A Novel Augmented Reality Ultrasound Framework using an RGB-D Camera and a 3D-printed Marker
\thanks{This work benefited from the European Unions Horizon 2020
research and innovation program under grant agreement n856950 (5G-TOURS
project). Also, it benefited from State aid managed by the National Research
Agency (FR) under the future investment program bearing the reference ANR-17-RHUS-0005 (FollowKnee project).} 
}

\author{
  Y. Zhou, G. Lelu, B. Labb\'{e}, G. Pasquier, P. Le Gargasson, A. Murienne and L. Launay \\
  Institute of Research and Technology b<>com \\
  Cesson-Sévigné, France \\
  \texttt{yitian.zhou@b-com.com}
}

\begin{document}
\maketitle

%%%%%%%%%%%%%%%%%%%%%%%%%%%%%%%%%%%%%%%%%%%%%%%%%%%%%%%%%%%%%%%%%%%%%%%%%%%%%%%%%%%%%%%%%
\begin{abstract}
\absdiv{Purpose}
 Ability to locate and track ultrasound images in the 3D operating space is of great benefit for multiple clinical applications. This is often accomplished by tracking the  probe  using  a precise but expensive optical  or electromagnetic  tracking  system. 
Our goal is to develop a simple and low cost augmented reality echography framework using a standard RGB-D Camera.

\absdiv{Methods}
A prototype system consisting of an Occipital Structure Core RGB-D camera, a specifically-designed 3D marker, and a fast point cloud registration algorithm FaVoR was developed and evaluated on an Ultrasonix ultrasound system. The probe was calibrated on a 3D-printed N-wire phantom using the software PLUS toolkit. The proposed calibration method is simplified, requiring no additional markers or sensors attached to the phantom. Also, a visualization software based on OpenGL was developed for the augmented reality application.

\absdiv{Results}
The calibrated probe was used to augment a real-world video in a simulated needle insertion scenario. The ultrasound images were rendered on the video, and visually-coherent results were observed. We evaluated the end-to-end accuracy of our AR US framework on localizing a cube of $5$ cm size. From our two experiments, the target pose localization error ranges from $5.6$ to $5.9$ mm and from  $-3.9^{\circ}$ to $4.2^{\circ}$. 

\absdiv{Conclusion}
We believe that with the potential democratization of RGB-D cameras integrated in mobile devices and AR glasses in the future, our prototype solution may facilitate the use of 3D freehand ultrasound in clinical routine.
Future  work  should include  a  more  rigorous  and  thorough  evaluation, by comparing the calibration accuracy with those obtained by commercial tracking solutions in both simulated and real medical scenarios.  
\end{abstract}

% keywords can be removed

%%%%%%%%%%%%%%%%%%%%%%%%%%%%%%%%%%%%%%%%%%%%%%%%%%%%%%%%%%%%%%%%%%%%%%%%%%%%%%%%%%%%%
\section{Introduction}
\label{intro}
Ultrasound (US) is a safe, portable and relatively cheap medical imaging modality. US machines produce real-time imaging of anatomical structures, providing essential information for many clinical practices such as disease diagnosis and intervention guidance. Despite recent developments in 3D transducer technology, 2D US is still the gold standard in clinical routine for its high image quality and low cost. 

In 2D US, images are acquired using a handheld probe, resulting in a series of 2D slices. Ability to locate and track those slices in the 3D operating space is of great benefit for multiple clinical applications, as it allows to register them to other signals or devices tracked in the same space. Examples include assisted needle biopsy, pre-operative or per-operative image fusion, and volumetric US reconstruction from a series of 2D slices. This is often accomplished by tracking the probe using an optical or electromagnetic (EM) tracking system. Given that the tracking system can only record the pose of the probe with respect to a world coordinate system, and not the image plane with respect to the world, an US Probe Calibration (UPC) procedure should be performed beforehand to obtain the spatial transform which maps coordinates in the US image plane to those of the probe~\cite{chen2016}.

A variety of calibration methods have been proposed in the literature~\cite{wen2020,Garcia2020,toews2017}. The reader is referred to~\cite{mercier2005} for a comprehensive review. Most of them operate by scanning a 3D phantom with known geometrical properties that may consist of points, wires or planes~\cite{wen2020,Garcia2020}. Among them, one popular option is to use the N-wire phantom where metal wires are designed to form \quotes{N} shapes~\cite{chen2006}. Recently, image-based auto-calibration was introduced in~\cite{toews2017}. Local image features are extracted from the patient's US data. Registration techniques are then used to establish spatial alignment between slices acquired from different probe positions, from which is further estimated the calibration matrix. The obtained accuracy was comparable with the state-of-the-art phantom-based methods~\cite{toews2017}. However, the authors claim that successful auto-calibrations require that the anatomy of interest contains distinctive, localizable image structure from which local features could be extracted~\cite{toews2017}, which remains one main limitation of the method.

An important component of the UPC and the associated Augmented Reality (AR) application is the underlying tracking system. Classical systems consist of central control units and optical/EM sensors~\cite{lasso2014}. Many commercial solutions exist, for example the NDI Polaris/Aurora and the CIVCO’s Ultra-Pro. Those systems ensure a high tracking precision. For example, the NDI Aurora has a localization precision of $\sim1.4$ mm~\cite{ndiwebsite}. The UPC is performed priorly, by attaching sensors to the phantom, to the probe and to the \textit{stylus} respectively~\cite{lasso2014}. The \textit{stylus} is an auxiliary tool used for finding the spatial relationship between the phantom and its attached sensor~\cite{lasso2014}. %Open-source softwares such as PLUS~\cite{lasso2014} are then utilized for the calibration.
Such solutions ensure usually a high calibration precision ($\sim1.1$ mm as reported in~\cite{Lange2011}). However, a costly, complex, and somewhat cumbersome system including a myriad of tools and accessories, as well as well-defined procedures, is required, which might hamper their utilization in clinical routine.

Recently, several groups~\cite{baba2016,busam2018} attempted to develop simpler and cheaper US tracking systems using standard RGB(-D) cameras. Both proposed \textit{inside-out} based tracking methods~\cite{busam2018}. The camera is fixed upon the probe and used actually as a sensor. Its pose is estimated from contextual information contained in the recorded video streams, using standard computer-vision techniques like the SLAM~\cite{busam2018}. %The camera acts actually as a sensor, and its pose is estimated from contextual information. %Both groups used PLUS~\cite{lasso2014} and the N-wire phantom for calibration. 
The authors stated that such \textit{inside-out} designs aimed to reduce the line-of-sight effect~\cite{busam2018}. Nonetheless, those methods usually require textured scenes and the presence of multiple fiducial markers in the Operating Room (OR)~\cite{busam2018}. In addition, RGB(-D) cameras of significant weight, as well as the associated power supply cable, need to be mounted on the probe. All those elements could have a non-negligible impact on the routine clinical environment. 

Several groups have gone a step further by proposing 3D freehand US reconstructions without any external tracking device~\cite{prevost2018}\cite{guo2020}. Deep learning techniques and inertial measurement units are combined to directly estimate the motion between successive frames from US data. Promising results are reported in~\cite{prevost2018}\cite{guo2020}. However, in~\cite{prevost2018}, the authors claim that no back-and-forth probe motion can be managed. In fact, the probe is supposed to move in a fixed direction, for instance, from proximal to distal, which could be troublesome in some clinical situations.  

In this paper, we propose a novel 2D UPC and the associated navigated US framework, combining an RGB-D camera, a 3D-printed marker and a previously-developed fast model-based 3D point cloud registration algorithm FaVoR~\cite{murienne2020}. Our method works in a more conventional \textit{outside-in} manner: real-time tracking of the marker by registering its virtual model (CAD mesh) with the 3D point cloud captured by the depth camera. Unlike the \textit{inside-out}-based solutions proposed in~\cite{baba2016,busam2018} where the camera's real-time pose is estimated from RGB(-D) information, the tracking via FaVoR is only based on depth information~\cite{murienne2020}. This is aimed at obtaining a stable solution despite the constraints related to the OR, such as the strong OR light which can sometimes saturate the RGB sensors. 

The main novelties are twofold: 1) FaVoR~\cite{murienne2020}, an RGB-D camera, and a specifically-designed 3D-printed marker dedicated for depth-based tracking are merged into an unified framework for AR US. 
    Depth cameras have been democratizing rapidly since several years, with the advantage of being much more affordable than commercial optical/EM tracking systems. We believe that this could significantly reduce the cost of navigated US in the future, therefore facilitating its clinical use; and 2) The probe calibration workflow is simplified, thanks to FaVoR which directly registers the virtual model of the phantom with the 3D representation of the world captured by the depth camera, therefore obviating the need to attach additional markers to the phantom, unlike most existing solutions~\cite{lasso2014}.

%%%%%%%%%%%%%%%%%%%%%%%%%%%%%%%%%%%%%%%%%%%%%%%%%%%%%%%%%%%%%%%%%%%%%%%%%%%%%%%%%%%%%%
\label{method}

\subsection{Ultrasound probe calibration}
\label{sec:upc}

\begin{figure}[!h]
  \centering
  \includegraphics[width=1.0\textwidth]
  {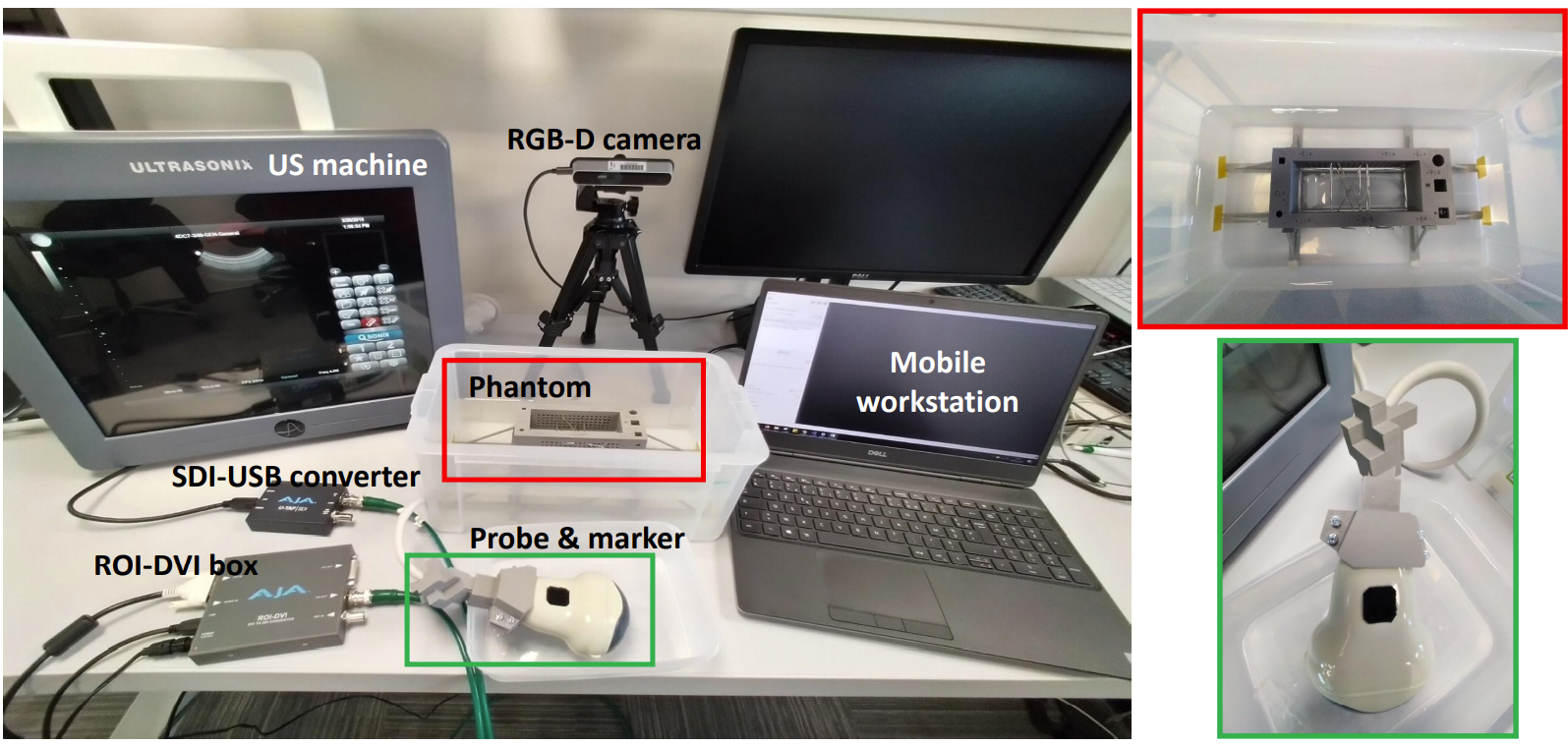}
\caption{The proposed ultrasound probe calibration setting which consists of an Ultrasonix US machine, an Occipital Structure Core RGB-D camera, a water tank containing the N-wire phantom, a mobile workstation and two US stream conversion boxes. The 3D-printed marker is rigidly fixed upon the probe. Zoomed views over the phantom and the probe are displayed in red and green boxes respectively. }
\label{fig:calib_setting} 
\end{figure}

As aforementioned, before deploying a navigated US system, the US image needs to be calibrated with respect to the probe marker. As shown in Fig.~\ref{fig:calib_setting}, our UPC system combines an RGB-D camera, an N-wire phantom, an US machine and a mobile workstation used for data processing. The marker is specifically designed to be empty inside so as to let pass through the cable. This specific design aims to avoid the occultation of the marker by the cable during the manipulation by clinicians.   

\subsubsection{Preparations}
\label{sec:upc:prep}
The RGB-D camera is installed on a tripod as shown in Fig.~\ref{fig:calib_setting}. We chose the Occipital Structure Core, due mainly to its high depth quality and framerate ($\sim$ $60$ Frame Per Second (FPS)). Its position and orientation are adjusted manually to be between 40 and 80 cm to the phantom, corresponding to the \textit{very short range} of the camera. The impact of this distance on the calibration accuracy will be studied later in Sect.~\ref{results}.

\paragraph{\textbf{Ultrasound image streaming}}

\begin{figure}[!h]
\centering
\includegraphics[width=0.95\textwidth]
 {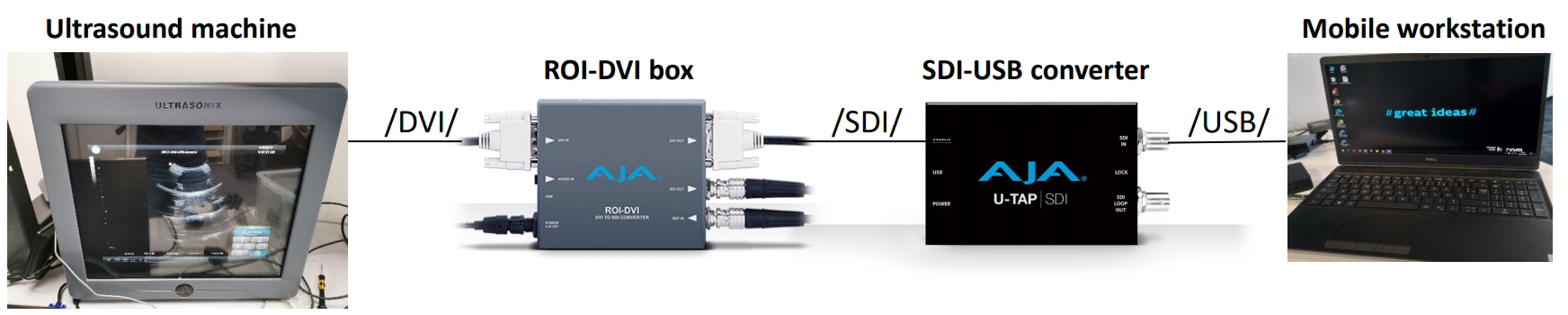}
\caption{Streaming of ultrasound images via an ROI-DVI box and a SDI-USB converter of the AJA company. }
\label{fig:streaming} 
\end{figure}

The Ultrasonix machine outputs the US images in DVI format. The goal is to feed the US stream to our mobile workstation via an USB port. A stream conversion system consisting of two AJA boxes was therefore designed. As illustrated in Fig.~\ref{fig:streaming}, a first ROI-DVI box is used to crop the original US images and output the stream in SDI format, which is further converted into USB format by the second SDI-USB converter. %Fig.~\ref{fig:streaming} illustrates the streaming conversion process.

\paragraph{\textbf{N-wire phantom}} 
The N-wire phantom \quotes{fCal-2.1} was used~\cite{lasso2014}. Additionally, a base was designed for fixing the N-wire phantom inside the water tank, since our method assumes that the phantom is not moving during the calibration. The phantom and the base are both 3D-printed. Nine metal wires of $1$ mm diameter were installed so as to form the \quotes{N} shape at three different depth levels.

\paragraph{\textbf{3D marker dedicated for depth cameras}} 
Fig.~\ref{fig:calib_setting} provides a zoomed view of the marker. The patent-pending marker results from a sophisticated combination of small cubes of size $1$ cm, forming an object of roughly $4$cm span. It was 3D-printed using plastic materials. This is aimed to obtain a sufficiently-textured while tiny and light object which can be easily localized by RGB-D cameras. In addition, a fixation system was 3D-printed in order to firmly fix the marker on the probe. In future applications, similar markers can also be attached to other devices, as the total weight of the marker and the fixation system is only about $30$g.

\subsubsection{Calibration workflow}
\label{sec:upc:workflow}

Our calibration method is based on the open-source software PLUS toolkit~\cite{lasso2014}. Following the official tutorials~\cite{lasso2014}, a new tracking device FaVoR~\cite{murienne2020} was defined and integrated into the \quotes{fCal} application~\cite{lasso2014}. Both the RGB-D camera and the US image stream are connected to the mobile workstation via USB 3.0 cables, and are managed by the \quotes{fCal} application. 

\begin{figure}[!h]
\centering
\includegraphics[width=1.0\textwidth]
 {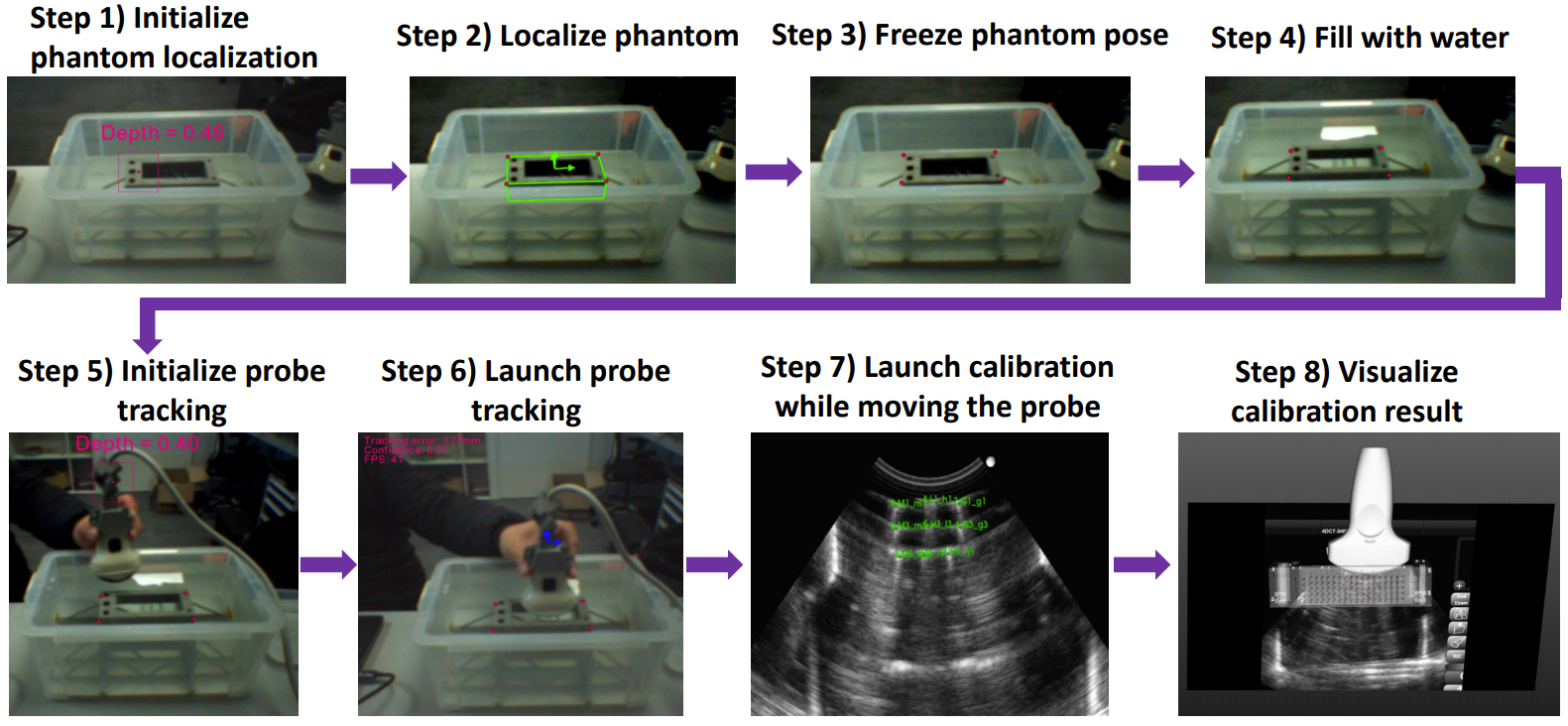}
\caption{The workflow of the proposed US probe calibration method. FaVoR~\cite{murienne2020} first localizes the phantom, and then tracks the probe marker. The poses are fed to PLUS~\cite{lasso2014} for calibration. }
\label{fig:calib_workflow} 
\end{figure}

The calibration framework comprises three key modules: phantom localization, probe tracking and PLUS calibration. Unlike most calibration procedures where additional markers/sensors need to be attached to the phantom for its localization~\cite{lasso2014}, we make full use of the generic nature of FaVoR and localize the N-wire phantom without any use of markers. For this, FaVoR takes the virtual model of the N-wire phantom as input and directly register it with the 3D point cloud captured by the depth camera. This generates the pose of the phantom in the camera's coordinate system. This pose is then frozen, leading to a static phantom pose (one simplification that we make is that the phantom is not moving during the calibration). Then, FaVoR is reparametrized with the marker's CAD model to estimate the pose of the probe marker in real time using the same RGB-D camera.

The probe is placed over the phantom for imaging the installed fiducial wires, producing US images with nine visible white spots. Those points are automatically segmented using functionnalities provided within PLUS~\cite{lasso2014}. Then, the three middle-wire points are matched to the groundtruth positions in the phantom~\cite{lasso2014}. Since both the phantom and the probe marker are localized by the same camera, spatial positions in the phantom space are easily mapped to 3D coordinates in the marker's space. This generates correspondences between the segmented image pixels and the spatial locations in the marker's space. From such correspondences, PLUS computes the calibration matrix using least-square fitting~\cite{lasso2014}.

The whole workflow is illustrated in Fig.~\ref{fig:calib_workflow}. The water tank is first half filled, exposing the phantom outside in the air. This is aimed to free the phantom localization from the optical refraction effect. Once localized, the pose of the phantom is frozen and stays constant during the calibration. More water is then added to the tank, covering entirely the phantom so that it can be visualized under US. FaVoR is then reconfigured to track the marker. Its pose, combined with that of the phantom and the streamed US images are then fed to PLUS \quotes{fCal} for calibration matrix estimation~\cite{lasso2014}. The technical details of the individual components are provided below. 

\paragraph{\textbf{Phantom localization}} FaVoR is parametrized with the virtual model (CAD mesh) of the fCal 2.1 phantom~\cite{lasso2014}. The goal is to register the virtual model with the 3D representation of the real world captured by the RGB-D camera. FaVoR requires a manual initialization. The RGB-D camera is adjusted manually to align the ROI center with a predefined zone of the phantom (Step 1 in Fig.~\ref{fig:calib_workflow}). The localization is then triggered, generating real-time poses of the phantom (Step 2 in Fig.~\ref{fig:calib_workflow}). Since we assume that neither the phantom nor the camera are allowed to move during the calibration, which is one constraint of our method, the pose of the phantom has constant values. As a result, the pose is frozen (Step 3 in Fig.~\ref{fig:calib_workflow}) so as to free FaVoR for the following tracking of the probe marker. 

\paragraph{\textbf{Probe tracking}} Similar to the localization of the phantom, FaVoR is reconfigured to register the virtual model of the marker with the 3D point cloud captured by the depth camera this time. The position of the probe is adjusted manually so as to align the ROI with a predefined zone on the marker (Step 5 in Fig.~\ref{fig:calib_workflow}). Once initialized, FaVoR tracks the marker at $\sim40$ FPS on the mobile workstation Intel(R) Core(TM) i7-10850H of 2.70GHz (Step 6 in Fig.~\ref{fig:calib_workflow}). 
 
\paragraph{\textbf{PLUS calibration}} After having properly positioned the probe over the phantom, the fiducial wires are visualized as white bright spots on US. They are segmented automatically using tools of PLUS which are based on mathematical morphology and thresholding. An example of the segmentation is shown in Fig.~\ref{fig:calib_workflow} (Step 7). The segmented spots are then matched to the nine wires by PLUS~\cite{lasso2014}. The pixel coordinates of the three middle-wire points in the US images are matched to 3D coordinates in the marker's coordinate system using the poses of the marker and the phantom~\cite{lasso2014}. PLUS then estimates the calibration matrix from these image-marker correspondences~\cite{lasso2014}. The calibrated US image and the phantom, as well as the installed wires, are visualized together under PLUS \quotes{fCal} (Step 8 in Fig.~\ref{fig:calib_workflow}). 

\subsection{Navigated US and augmented real-world video}
\label{sec:arus}
The calibrated probe and the RGB-D camera-based tracking system are deployed to augment a real-world scenario with US image information. A visualization software based on OpenGL has been developed for rendering an US image within a standard RGB image. It takes as input the marker's pose estimated by FaVoR, the calibration matrix, and the intrinsic parameters of the color sensor of the RGB-D camera, and renders both the marker and the US image in real-time. %An example of our AR US will be shown later in Fig.~\ref{fig:visu_result_blueprint} in Sect.~\ref{results}. 

%%%%%%%%%%%%%%%%%%%%%%%%%%%%%%%%%%%%%%%%%%%%%%%%%%%%%%%%%%%%%%%%%%%%%%%%%%%%%%%%%%%

\section{Result}
\label{results}
\subsection{Evaluation within PLUS}
\subsubsection{Tuning the camera-to-phantom distance}
In our case, the distance between the RGB-D camera and the top of the phantom is a key variable that could impact the calibration accuracy, and therefore the precision of the AR US application. As a result, we performed the calibration at five different distances ranging from $40$ to $80$ cm. At each distance, five calibrations were repeated. The magnitudes of the calibration errors on N-wire phantom were listed in Tab.~\ref{tab:1}. We observe that the Occipital Structure Core camera achieved its best accuracy at $50$ cm, corresponding to an average calibration error of $2.6$ mm. The calibration errors were computed by PLUS on additional US images acquired on the N-wire phantom~\cite{lasso2014}. The wires were segmented by PLUS and reprojected to the marker's space using the calibration matrix. Those points are then transformed to the phantom's space since both the phantom and the marker are localized by the same camera. Distances between those points and the groundtruth were computed and then averaged. At $40$ cm, bigger errors were obtained. This is because that while scanning the probe, the distance between the marker and the camera could be below $30$ cm, which is beyond the operating range of the depth camera.

\begin{table}[!h]
\centering
\caption{Evolution of the calibration error over the camera-phantom distance.}
\label{tab:1}       % Give a unique label
\begin{tabular}{l|ccccc}
\hline\noalign{\smallskip}
Distance (cm) & 40  & 50 & 60 & 70 & 80  \\
\noalign{\smallskip}\hline\noalign{\smallskip}
Calib. Err. (mm) & $30.1\pm 20.0$  
& 
$2.6 \pm 0.4$ 
& 
$3.4 \pm 0.5$ 
&  
$3.3 \pm 0.7$   
&  
$4.0 \pm 0.6$ \\
\noalign{\smallskip}\hline
\end{tabular}
\end{table}

\subsubsection{Visualizing the calibration result on N-wire phantom}
\begin{figure}[!h]
     \centering
         \centering
         \includegraphics[width=0.6\textwidth]{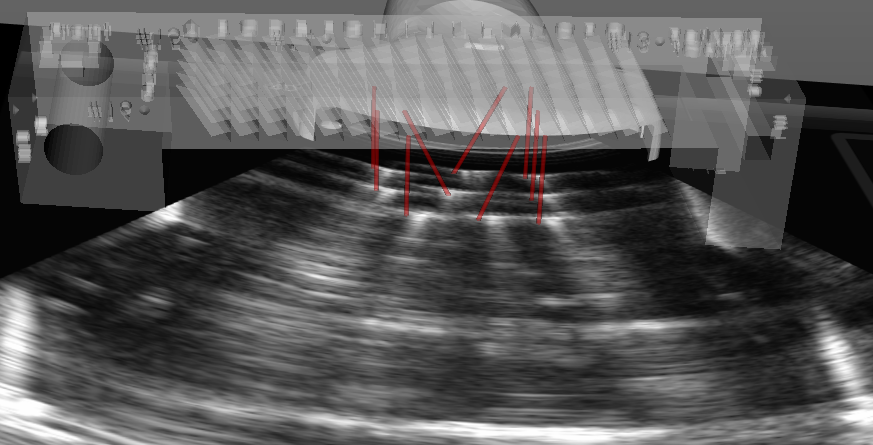}
         \caption{US image calibrated with the N-wire phantom visualized in \quotes{fCal} of PLUS~\cite{lasso2014} (red lines show the groundtruth wires)}
    \label{fig:visu_result_plus}
\end{figure}
Fig.~\ref{fig:visu_result_plus} shows an example of the calibrated image under PLUS \quotes{fCal} at the optimal camera-to-phantom distance of $50$ cm. The groundtruth wires are shown as red lines, and we can observe that they are aligned with the white spots in the US images (metal wires are visualized in US as higher intensity pixels). 

\subsection{Simulation of needle insertion on a BluePrint phantom}
\begin{figure}[!h]
     \centering
    \includegraphics[width=0.85\textwidth]{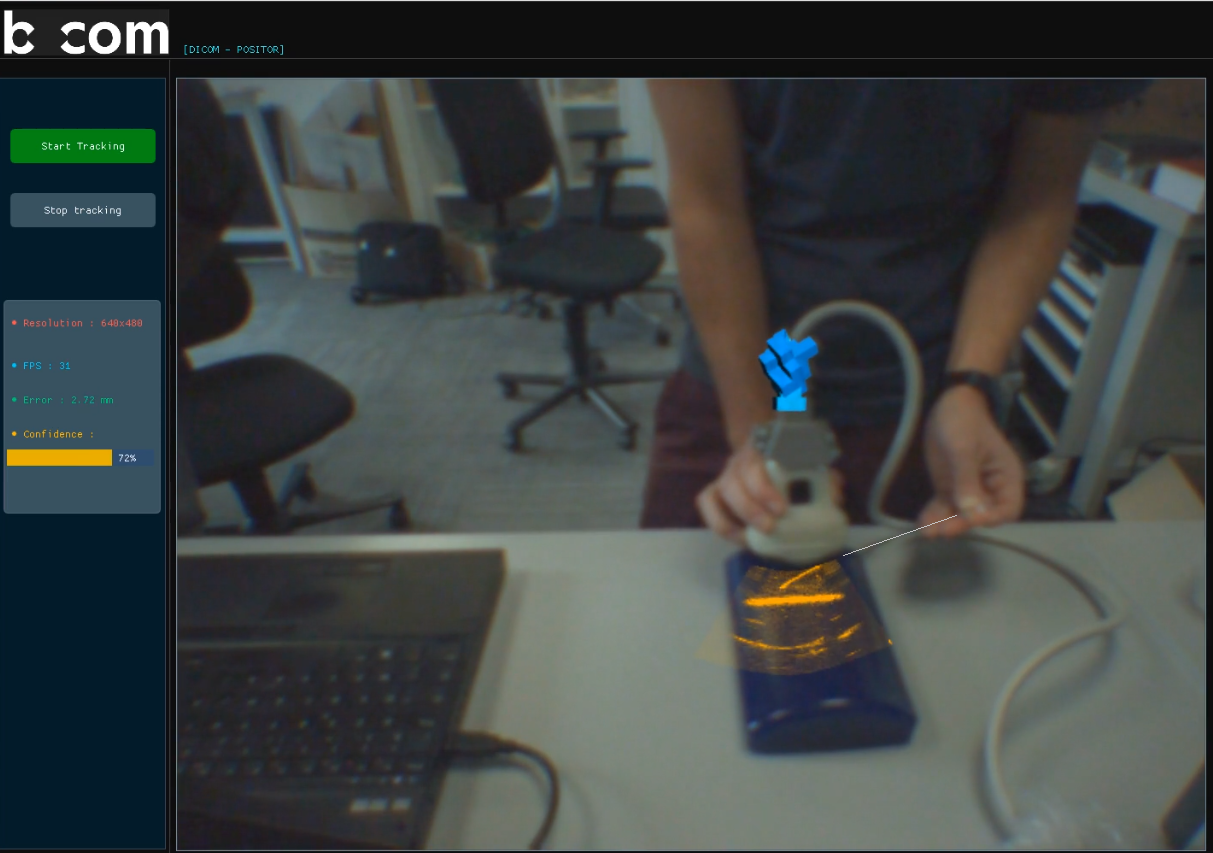}
    \caption{Augmented ultrasound simulating needle insertion in real-world video (the tracked marker is rendered in blue and the calibrated US image is displayed on the RGB image)}
    \label{fig:visu_result_blueprint}
\end{figure}
Fig.~\ref{fig:visu_result_blueprint} shows the deployment of the calibrated probe for augmenting a real-world video on a BluePrint phantom. A needle insertion procedure was simulated. The US images are rendered using a proper color map which highlights pixels with higher intensities in bright yellow. As we can observe in Fig.~\ref{fig:visu_result_blueprint}, the part of the needle which is originally invisible in the RGB image, since it is inside the phantom, is displayed in the US image on the video. A spatial continuity is observed between the needle part which is outside the phantom (highlighted for better visualization as the color sensor of the Occipital Structure Core camera has a rather limited spatial resolution) and the inside part revealed by US. 

\subsection{Evaluation of the localization accuracy of a cube using our AR US framework}
\begin{figure}[!h]
     \centering
     \begin{subfigure}[b]{0.48\textwidth}
         \centering
         \includegraphics[width=\textwidth]{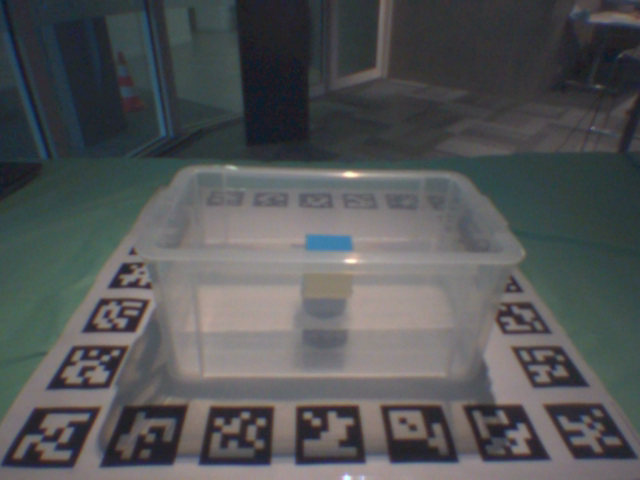}
         \caption{A cube is fixed inside a water tank which is carefully placed at a predefined position with respect to the ArUco markers}
         \label{fig:cube_inside_markers}
     \end{subfigure}
     \hfill
     \begin{subfigure}[b]{0.48\textwidth}
         \centering
         \includegraphics[width=\textwidth]{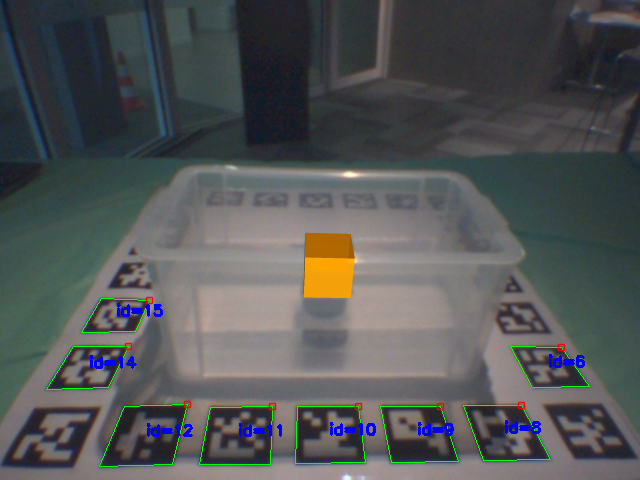}
         \caption{The groundtruth pose of the cube in the camera space is estimated using RGB-D camera-based pose estimation techniques}
         \label{fig:cube_inside_markers_gt}
     \end{subfigure}
    \caption{Generation of the groundtruth of the cube's pose in the RGB-D camera's space using ArUco markers.}
    \label{fig:visu_result}
\end{figure}

We aim to further evaluate the end-to-end accuracy of our AR US framework. A cube of size $5$ cm was used. Similar to the method used in~\cite{hinterstoisser2012}, the cube was placed at a precise position in the middle of a planar board with markers attached to it, as shown in Fig.~\ref{fig:cube_inside_markers}. The board defines a world Coordinate System (CS) with its origin at the center. All the marker corners' positions are known. From the RGB images, part of the markers can be localized using OpenCV~\cite{hinterstoisser2012}, as is displayed in Fig.~\ref{fig:cube_inside_markers_gt}. The localized 2D pixels are further unprojected to the 3D CS of the camera, using the depth map and the intrinsic parameters of the color sensor. This generates 3D-to-3D correspondences between the camera and world CS. The pose of the cube is then computed using the technique introduced in~\cite{gonzalez2021}. In this way, we obtain the groundtruth position of the cube, as displayed in Fig.~\ref{fig:cube_inside_markers_gt}.

\begin{figure}[!h]
     \centering
       \includegraphics[width=\textwidth]{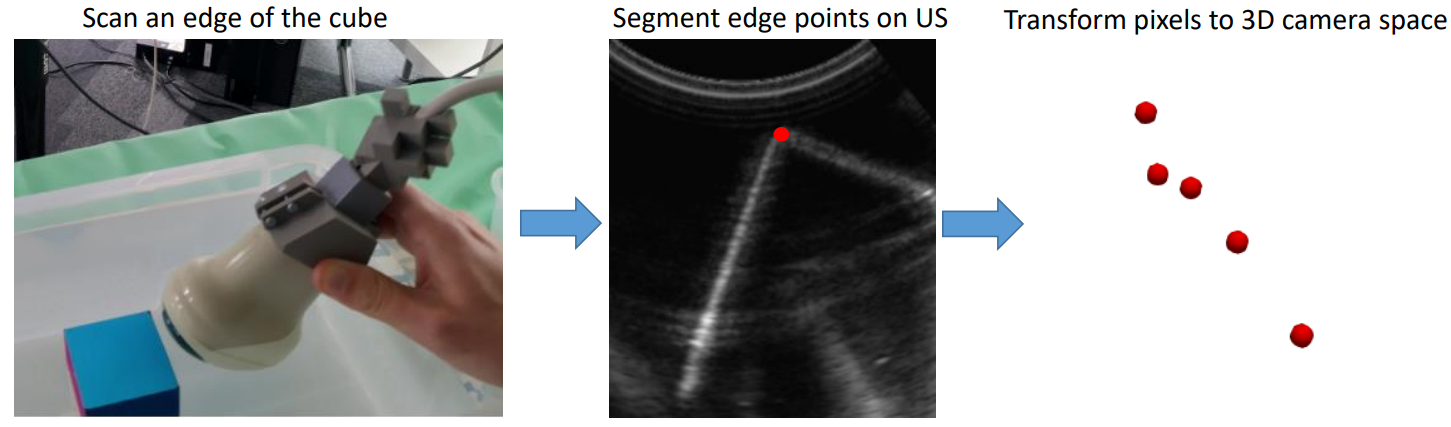}
         \caption{Scan the cube edges using the proposed AR US system. The segmented US pixels are further unprojected to 3D points in the RGB-D camera's space, through the calibration matrix and probe marker's poses. }
         \label{fig:scan_cube}
\end{figure}

\begin{figure}
    \centering
    \begin{subfigure}[b]{0.25\textwidth}
         \centering
         \includegraphics[width=\textwidth]{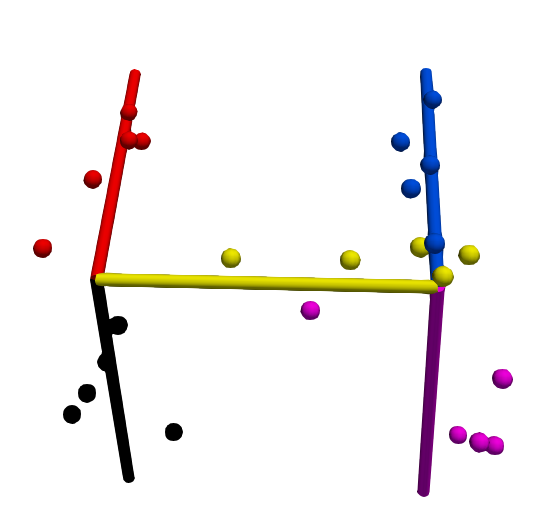}
         \caption{localized cube edge points (view point I)}
         \label{fig:visu_eva_cube_a}
     \end{subfigure}
     \hfill
     \begin{subfigure}[b]{0.25\textwidth}
         \centering
         \includegraphics[width=\textwidth]{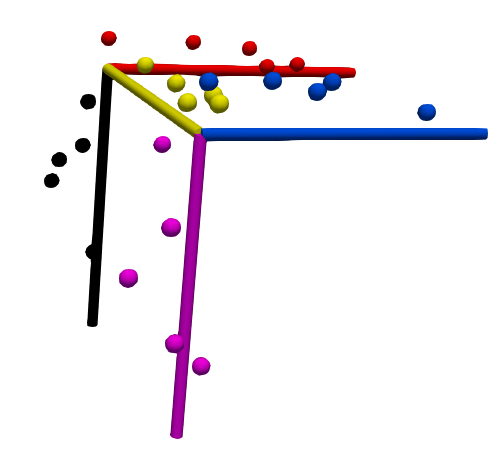}
         \caption{localized edge points (view point II)}
         \label{fig:visu_eva_cube_b}
     \end{subfigure}
     \hfill
     \begin{subfigure}[b]{0.18\textwidth}
         \centering
         \includegraphics[width=\textwidth]{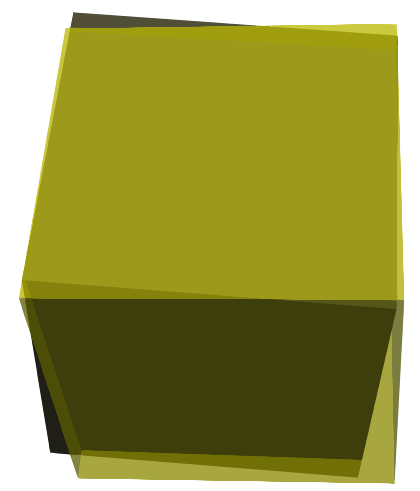}
         \caption{cube fitted from the localized edge points (view point I)}
         \label{fig:visu_eva_cube_c}
     \end{subfigure}
     \hfill
     \begin{subfigure}[b]{0.19\textwidth}
         \centering
         \includegraphics[width=\textwidth]{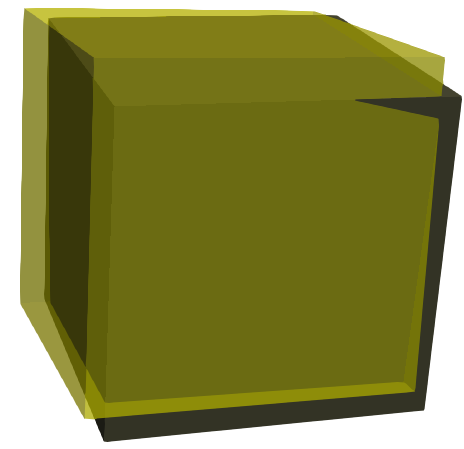}
         \caption{cube fitted from the localized edge points (view point II)}
         \label{fig:visu_eva_cube_d}
     \end{subfigure}
    \caption{Evaluation of the cube localization accuracy obtained by our AR US method. In (a) and (b), solid lines show groundtruth cube edges and the localization results are displayed as small spheres of the same color. In (c) and (d), groundtruth cube is displayed in black color, and the fitted one in yellow. The distance between the two cube centers is $5.9$ mm, and the three Euler angles offsets are $0.8^{\circ}$, $-3.9^{\circ}$ and $4.2^{\circ}$. }
    \label{fig:visu_eva_cube}
\end{figure}

Now that the groundtruth is available, we are able to evaluate the accuracy of our AR US system quantitatively. For this, as illustrated in Fig.~\ref{fig:scan_cube}, five edges of the cube are scanned, which is enough to reconstruct the cube. Pixels belonging to the edge are segmented and transformed to the probe marker's CS via the calibration matrix. Since the marker is tracked, those points are further mapped to the camera's CS where the groundtruth is available (Fig.~\ref{fig:visu_result}). Fig.~\ref{fig:visu_eva_cube_a} and \ref{fig:visu_eva_cube_b} show the groundtruth cube edges and the corresponding localization points generated using the method introduced in Fig.~\ref{fig:scan_cube}. Iterative Closest Point (ICP) was then used to compute the rigid transformation between the groundtruth edges and the localized points. The transformation was applied to the groundtruth cube, yielding the reconstructed cube as displayed in Fig~\ref{fig:visu_eva_cube_a} and \ref{fig:visu_eva_cube_b}. The ICP registration has a residue of $5.4$ mm average distance. An additional experiment was performed, and the obtained accuracies from the two experiments are reported in Tab.~\ref{tab:2}. %The distance between the cube centers is $5.9$ mm, and the three Euler angles are $0.8^{\circ}$, $-3.9^{\circ}$ and $4.2^{\circ}$.   
Note that here the level of accuracy corresponds to the whole AR US method, which combines the probe calibration error, the probe tracking error via FaVoR, potential manual US image segmentation errors and finally geometric distortions in  US images. This could partially explain why the error magnitude is higher than the tracking error reported in~\cite{murienne2020}.

\begin{table}[!h]
\centering
\caption{The level of accuracy obtained by our augmented reality ultrasound framework for localizing a cube of $5$ cm size.}
\label{tab:2}       % Give a unique label
\begin{tabular}{l|cccc}
\hline\noalign{\smallskip}
 &  \begin{tabular}{@{}c@{}}ICP registration residue \\ (mean dist. in mm)  \end{tabular}  & 
 \begin{tabular}{@{}c@{}}Cube center offset \\ (mm)  \end{tabular} & 
  \begin{tabular}{@{}c@{}}Euler angles offset \\ (degree)  \end{tabular}  \\
\noalign{\smallskip}\hline\noalign{\smallskip}
Experiment n$^{\circ}$ 1 & 5.4 & 5.9  &  \{0.8, -3.9, 4.2\}   \\
Experiment n$^{\circ}$ 2 & 4.5 & 5.6  &  \{1.8, 1.4, -2.5\}   \\
\noalign{\smallskip}\hline
\end{tabular}
\end{table}

%%%%%%%%%%%%%%%%%%%%%%%%%%%%%%%%%%%%%%%%%%%%%%%%%%%%%%%%%%%%%%%%%%%%%%%%%%%%%%%%%%%%%%
\section{Conclusion}
A new simple and low cost AR US framework including essentially an RGB-D Camera, a 3D-printed marker, and a fast point-cloud registration algorithm FaVoR was developed and evaluated on an Ultrasonix US system. The probe calibration was performed using the N-wire phantom and the software PLUS~\cite{lasso2014}. Preliminary results using the Occipital Structure Core showed a mean fiducial-based calibration error of $2.6$ mm generated within the PLUS application, corresponding to the camera to phantom distance of around $50$ cm. %This is the first time that RGB-D camera was used for US probe calibration in a conventional \quotes{outside-in} manner. 
The calibrated probe was then used to augment a real-world video in a simulated needle insertion scenario. Visually-coherent results were observed by showing a spatial continuity between the part of the needle outside the phantom visible in RGB images and the remaining part revealed by US. Furthermore, we evaluated the the end-to-end accuracy of our AR US framework on localizing a cube of $5$ cm size. From our two experiments, the target pose localization error ranges from $5.6$ to $5.9$ mm and from  $-3.9^{\circ}$ to $4.2^{\circ}$.

One limitation of our method is that during the calibration, neither the camera nor the water tank upon which the N-wire phantom is fixed are allowed to move. 
Future work should include a more comprehensive validation of the proposed method, through quantitative evaluations in both simulated and real medical scenarios. The planar board could be 3D-printed together with the evaluation object in order to minimize all potential errors in groundtruth generation due to misalignment between the object and the board. More complex 3D shapes, for instance anatomical organ models, should also be used for the evaluation. Also, both the probe tracking and the calibration should be further optimized by using a marker mesh of a higher spatial resolution (currently downsampled to $3$ mm for higher FPS $\sim50$) in order to improve the accuracy. A thorough comparison with existing AR US solutions relying on commercial tracking system in terms of both accuracy and computational time will also be beneficial. Besides, currently we used the Occipital Structure Core, a comprehensive comparison of different commercial RGB-D cameras, regarding both the depth and color sensors, will be useful. In future extensions, the system shall also support the use of multiple 3D markers in order to track more than one device. We believe that with the potential future democratization of RGB-D cameras integrated in both mobile devices and AR glasses, our prototype solution may facilitate the use of 3D freehand ultrasound in clinical practices.

%\section{Conclusion}
%Your conclusion here

%\section*{Acknowledgments}
%This was was supported in part by......

%Bibliography
\bibliographystyle{unsrt}  

%\bibliography{references}  

\end{document}